\documentclass[final,3p,times,a4paper]{elsarticle}

\usepackage{graphicx}
\usepackage{amssymb}
\usepackage{amsmath}
\usepackage{hyperref}
\usepackage[usenames]{color}

\journal{Computer Physics Communications}

\begin{document}

\begin{frontmatter}

\title{OpenMP, OpenMP/MPI, and CUDA/MPI C programs for solving the time-dependent dipolar Gross-Pitaevskii equation}

\author[scl]{Vladimir Lon\v{c}ar\corref{author}}
\ead{vladimir.loncar@ipb.ac.rs}

\author[ust,ift]{Luis E. Young-S.}
\ead{luisevery@gmail.com}

\author[dmi]{Srdjan \v{S}krbi\'{c}}
\ead{srdjan.skrbic@dmi.uns.ac.rs}

\author[bdu]{Paulsamy Muruganandam}
\ead{anand@cnld.bdu.ac.in}

\author[ift]{Sadhan K. Adhikari}
\ead{adhikari@ift.unesp.br}

\author[scl]{Antun Bala\v{z}}
\ead{antun.balaz@ipb.ac.rs}

\cortext[author] {Corresponding author.}
\address[scl]{Scientific Computing Laboratory, Center for the Study of Complex Systems, Institute of Physics Belgrade,\\ University of Belgrade, Pregrevica 118, 11080 Belgrade, Serbia}
\address[ust]{Departamento de Ciencias B\'{a}sicas, Universidad Santo Tom\'{a}s, 150001 Tunja, Boyac\'{a}, Colombia}
\address[ift]{Instituto de F\'{\i}sica Te\'{o}rica, UNESP -- Universidade Estadual Paulista, 01.140-70 S\~{a}o Paulo, S\~{a}o Paulo, Brazil}
\address[dmi]{Department of Mathematics and Informatics, Faculty of Sciences, University of Novi Sad, Trg Dositeja Obradovi\' ca 4, 21000 Novi Sad, Serbia}
\address[bdu]{School of Physics, Bharathidasan University, Palkalaiperur Campus, Tiruchirappalli -- 620024, Tamil Nadu, India}

\vspace*{-15mm}
\begin{abstract}
We present new versions of the previously published C and CUDA programs for solving the dipolar Gross-Pitaevskii equation in one, two, and three spatial dimensions, which calculate stationary and non-stationary solutions by propagation in imaginary or real time. Presented programs are improved and parallelized versions of previous programs, divided into three packages according to the type of parallelization. First package contains improved and threaded version of sequential C programs using OpenMP. Second package additionally parallelizes three-dimensional variants of the OpenMP programs using MPI, allowing them to be run on distributed-memory systems. Finally, previous three-dimensional CUDA-parallelized programs are further parallelized using MPI, similarly as the OpenMP programs. We also present speedup test results obtained using new versions of programs in comparison with the previous sequential C and parallel CUDA programs. The improvements to the sequential version yield a speedup of 1.1 to 1.9, depending on the program. OpenMP parallelization yields further speedup of 2 to 12 on a 16-core workstation, while OpenMP/MPI version demonstrates a speedup of 11.5 to 16.5 on a computer cluster with 32 nodes used. CUDA/MPI version shows a speedup of 9 to 10 on a computer cluster with 32 nodes.
\end{abstract}


\end{frontmatter}

\begin{small}
\noindent
{\bf New version program summary} \vspace*{1mm}\\
\noindent
{\em Program Title:} DBEC-GP-OMP-CUDA-MPI: (1) DBEC-GP-OMP package: (i) imag1dX-th, (ii) imag1dZ-th, (iii) imag2dXY-th, (iv) imag2dXZ-th, (v) imag3d-th, (vi) real1dX-th, (vii) real1dZ-th, (viii) real2dXY-th, (ix) real2dXZ-th, (x) real3d-th; (2) DBEC-GP-MPI package: (i) imag3d-mpi, (ii) real3d-mpi; (3) DBEC-GP-MPI-CUDA package: (i) imag3d-mpicuda, (ii) real3d-mpicuda.\vspace*{1mm}\\
{\em Program Files doi:} \href{http://dx.doi.org/10.17632/j3z9z379m8.1}{http://dx.doi.org/10.17632/j3z9z379m8.1}\vspace*{1mm}\\
{\em Licensing provisions:} Apache License 2.0\vspace*{1mm}\\
{\em Programming language:} OpenMP C; CUDA C. \vspace*{1mm}\\
{\em Computer:} DBEC-GP-OMP runs on any multi-core personal computer or workstation with an OpenMP-capable C compiler and FFTW3 library installed. MPI versions are intended for a computer cluster with a recent MPI implementation installed. Additionally, DBEC-GP-MPI-CUDA requires CUDA-aware MPI implementation installed, as well as that a computer or a cluster has Nvidia GPU with Compute Capability 2.0 or higher, with CUDA toolkit (minimum version 7.5) installed.\vspace*{1mm} \\
{\em Number of processors used:} All available CPU cores on the executing computer for OpenMP version, all available CPU cores across all cluster nodes used for OpenMP/MPI version, and all available Nvidia GPUs across all cluster nodes used for CUDA/MPI version. \vspace*{1mm}\\
{\em Journal reference of previous version:} Comput. Phys. Commun. \textbf{195} (2015) 117; {\it ibid.} \textbf{200} (2016) 406.\\
{\em Does the new version supersede the previous version?:} Not completely. OpenMP version does supersede previous AEWL\_v1\_0 version, while MPI versions do not supersede previous versions and are meant for execution on computer clusters and multi-GPU workstations.

\noindent\\
{\em Nature of problem:}
These programs are designed to solve the time-dependent nonlinear partial differential Gross-Pitaevskii (GP) equation with contact and dipolar interaction in a harmonic anisotropic trap. The GP equation describes the properties of a dilute trapped Bose-Einstein condensate. OpenMP package contains programs for solving the GP equation in one, two, and three spatial dimensions, while MPI packages contain only three-dimensional programs, which are computationally intensive or memory demanding enough to require such level of parallelization.

\noindent\\
{\em Solution method:}
The time-dependent GP equation is solved by the split-step Crank-Nicolson method by discretizing in space and time. The discretized equation is then solved by propagation, in either imaginary or real time, over small time steps. The contribution of the dipolar interaction is evaluated by a Fourier transformation to momentum space using a convolution theorem. MPI parallelization is done using the domain decomposition. The method yields the solution of stationary and/or non-stationary problems.

\noindent\\
{\em Reasons for the new version:}
Previously published C and Fortran programs \cite{dbec2015} for solving the dipolar GP equation are sequential in nature and do not exploit the multiple cores or CPUs found in typical modern computers. A parallel implementation exists, using Nvidia CUDA \cite{cuda2016}, and both versions are already used within the ultra-cold atoms community \cite{uca}. However, CUDA version requires special hardware, which limits its usability. Furthermore, many researchers have access to high performance computer clusters, which could be used to either further speed up the computation, or to work with problems which cannot fit into a memory of a single computer. In light of these observations, we have parallelized all programs using OpenMP, and then extended the parallelization of three-dimensional programs using MPI to distributed-memory clusters. Since the CUDA implementation uses the same algorithm, and thus has the same structure and flow, we have applied the same data distribution scheme to provide the distributed-memory CUDA/MPI implementation of three-dimensional programs.

\noindent\\
{\em Summary of revisions:}

\noindent\\
{\bf Package DBEC-GP-OMP:}
Previous serial C programs \cite{dbec2015} are here improved and then parallelized using OpenMP (package DBEC-GP-OMP). The main improvement consists of switching to real-to-complex (R2C) Fourier transform, which is possible due to the fact that input of the transform is purely real. In this case the result of the transform has Hermitian symmetry, where one half of the values are complex conjugates of the other half. The fast Fourier transformation (FFT) libraries we use can exploit this to compute the result faster, using half the memory. 

To parallelize the programs, we have used OpenMP with the same approach as described in \cite{bec2012}, and extended the parallelization routines to include the computation of the dipolar term. The FFT, used in computation of the dipolar term, was also parallelized in a straightforward manner, by using the built-in support for OpenMP in FFTW3 library \cite{FFTW3}. With the introduction of multiple threads memory usage has increased, driven by the need to have some variables private to each thread. To reduce the memory consumed, we resorted to using techniques similar to the ones used in our CUDA implementation \cite{cuda2016}, i.e., we have reduced the memory required for FFT by exploiting the aforementioned R2C FFT, and reused the memory with pointer aliases whenever possible.

\noindent\\
{\bf Package DBEC-GP-MPI:}
Next step in the parallelization (package DBEC-GP-MPI) was to extend the programs to run on distributed-memory systems, i.e., on computer clusters using domain decomposition with MPI programming paradigm. We chose to use the newly-implemented threaded versions of the programs as the starting point. Alternatively, we could have used serial versions, and attempt a pure MPI parallelization, however we have found that OpenMP-parallelized routines better exploit the data locality and thus outperform the pure MPI implementation. Therefore, our OpenMP/MPI-parallelized programs are intended to run one MPI process per cluster node, and each process would spawn the OpenMP threads as needed on its cluster node. Note that this is not a requirement, and users may run more than one MPI process per node, but we advise against it due to performance reasons. With the suggested execution strategy (one MPI process per cluster node, each spawning as many threads as CPU cores available), OpenMP threads perform most of the computation, and MPI is used for data exchanges between processes.

There are numerous ways to distribute the data between MPI processes, and we decided to use a simple one-dimensional data distribution, also known as slab decomposition. Data is distributed along the first (slowest changing) dimension, which corresponds to NX spatial dimension in our programs (see Fig.~\ref{fig1}). Each process is assigned a different portion of the NX dimension, and contains the entire NY and NZ spatial dimensions locally. This allows each process to perform computation on those two dimensions in the same way as before, without any data exchanges. In case the computation requires whole NX dimension to be local to each process, we transpose the data, and after the computation, we transpose the data back.

\begin{figure}[!t]
\centering
\includegraphics[width=8.5cm]{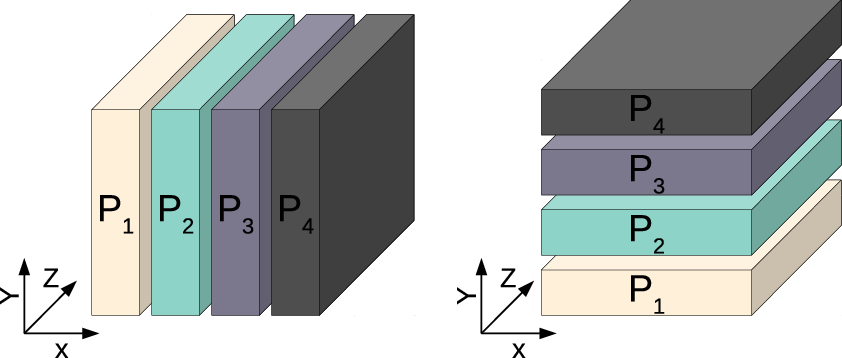}
\caption{Illustration of data distribution between MPI processes. On the left, the data are distributed along the NX dimension, while on the right the same data are redistributed along the NY dimension.}
\label{fig1}
\end{figure}

Transpose routine can be implemented in many ways using MPI, most commonly using \texttt{MPI\_Alltoall} function, or using transpose routines from external libraries, like FFTW3 \cite{FFTW3} or 2DECOMP\&FFT \cite{2decomp}. Since we already rely on FFTW3 library for FFT, we have utilized its dedicated transpose interface to perform the necessary transformations. To speed up transpose operation, we do not perform full transposition of data, but rather leave it \textit{locally} transposed. That is, we transform from local\_NX $\times$ NY $\times$ NZ, stored in row-major order, to NX $\times$ local\_NY $\times$ NZ in row-major order (where local\_NX = NX / number\_of\_processes, and equivalently for local\_NY). This approach has an additional benefit that we do not have to make significant changes in the way array elements are processed, and in most cases we only have to adjust the loop limit of the non-local dimension.

\noindent\\
{\bf Package DBEC-GP-MPI-CUDA:}
The aforementioned data distribution scheme can be also applied to the CUDA version of programs \cite{cuda2016}. However, there is no support for CUDA in FFTW3, and cuFFT (used in CUDA programs for FFT) does not provide equivalent MPI or transpose interface. Instead, we developed our own transpose routines, and used them in FFT computation. One example of manual implementation of transpose routines is shown in Ref.~\cite{becmpi2016}, and while we could readily use the same code, we wanted to have the same result as when using FFTW3. To achieve this, we use the same basic principle as in Ref.~\cite{becmpi2016}, first we create a custom MPI data type that maps to portions of the data to be exchanged, followed by an all-to-all communication to exchange the data between processes, see Fig.~\ref{fig2} for details.

\begin{figure}[!t]
\centering
\includegraphics[width=14cm]{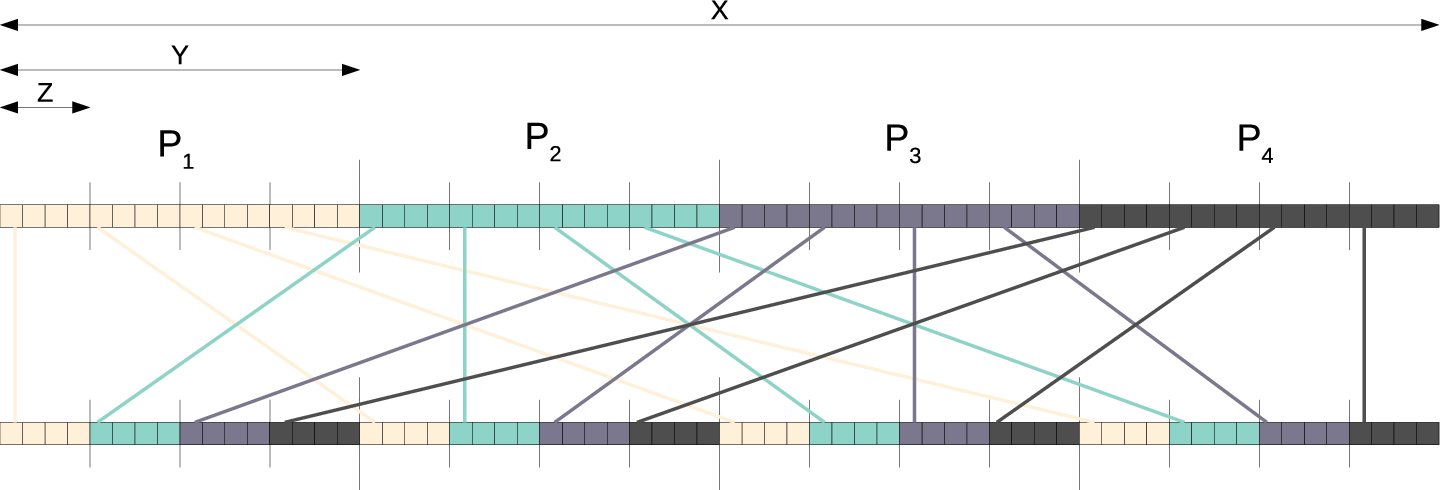}
\caption{Example of a transpose routine of a $4 \times 4 \times 4$ data between four MPI processes. Initially, all processes have $1/4$ of the NX dimension, and whole NY and NZ dimensions. After transposing, each process has full NX and NZ dimensions, and $1/4$ of the NY dimension. }
\label{fig2}
\end{figure}

The implemented transpose routines are also used to compute a distributed-memory FFT, performed over all MPI processes. To divide the computation of a multidimensional FFT, in our case three-dimensional, we use a well-known row-column algorithm. The basic idea of the algorithm is perhaps best explained on a two-dimensional FFT of $N \times M$ data, stored in row-major order, illustrated in Fig.~\ref{fig3}. First the $N$ one-dimensional FFTs of length $M$ are performed (along the row of data), followed by a transpose, after which data are stored as $M \times N$ in row-major format. Now $M$ FFTs of length $N$ can be performed along what used to be a column of original data, but are stored as rows after transposing. Finally, an optional transpose can be performed to return the data in their original $N \times M$ form. In three dimensions, we can perform a two-dimensional FFT, transpose the data, and perform the FFT along the third dimension. This algorithm can be easily adapted for distributed memory systems. We use advanced cuFFT interface for local computation of FFT, and use our transpose routine to redistribute the data.

Note that DBEC-GP-MPI-CUDA programs can be easily modified to work on a single workstation with multiple GPU cards, or a computer cluster with multiple GPU cards per node. In that case, for each GPU card a separate MPI process should be launched and the programs should be modified to assign a separate GPU card for processes on the same cluster node.

\begin{figure}[!t]
\centering
\includegraphics[width=13cm]{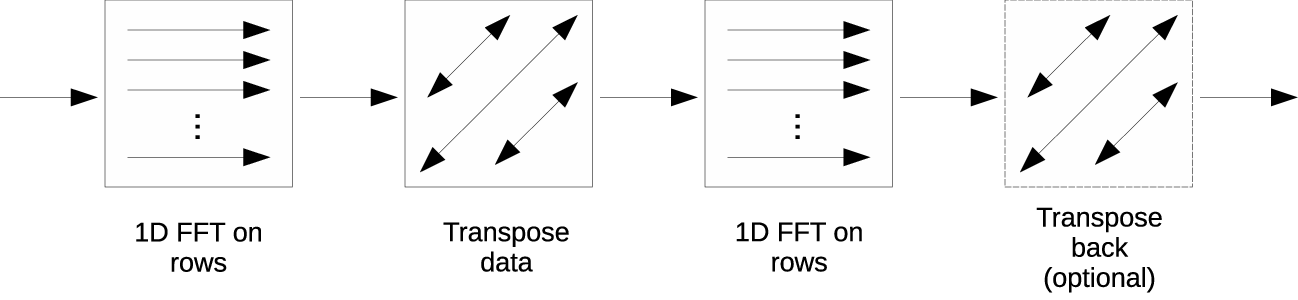}
\caption{Illustration of four stages of row-column FFT algorithm. The last transpose operation may be omitted, and often yields better performance. }
\label{fig3}
\end{figure}

\noindent\\
{\bf MPI output format:}
Given that the distributed memory versions of the programs can be used for much larger grid sizes, the output they produce (i.e., the density profiles) can be much larger and difficult to handle. To alleviate this problem somewhat, we have switched to a binary output instead of the textual. This allowed us to reduce the size of files, while still retaining precision. All MPI processes will write the output to the same file, at the corresponding offset, relieving the user of the task of combining the files. The binary output can be subsequently converted to textual, for example by using \texttt{hexdump} command on UNIX-like systems. We have developed a simple script which converts the output from binary to textual format and included it in the software package.

\noindent\\
{\bf Testing results:}
We have tested all programs on the PARADOX supercomputing facility at the Scientific Computing Laboratory of the Institute of Physics Belgrade. Nodes used for testing had two Intel Xeon E5-2670 CPUs (with a total of $2\times8=16$ CPU cores) with 32 GB of RAM and one Nvidia Tesla M2090 GPU with 6 GB of RAM, each connected by Infiniband QDR interconnect. The presented results are obtained for arbitrary grid sizes, which are not tailored to maximize performance of the programs. We also stress that execution times and speedups reported here are calculated for critical parallelized parts of the programs performing iterations over imaginary or real time steps, and they exclude time spent on initialization (threads initialization, MPI environment, allocation/deallocation of memory, creating/destroying FFTW plans, I/O operations). As a part of its output, each program separately prints initialization time and time spent on iterations for GP propagation. The latter time is used to calculate a speedup, as a speedup obtained this way does not depend on the number of iterations and is more useful for large numbers of iterations.

\begin{table}[!b]
\caption{Wall-clock execution times of DBEC-GP-OMP programs compiled with Intel's icc compiler, compared to the execution times of previously published serial versions. The execution times given here are for 1000 iterations (in seconds, excluding initialization and input/output operations, as reported by each program) with grid sizes: $10^5$ for 1d programs, $10^4\times 10^4$ for 2d programs, and $480\times 480 \times 480$ for 3d programs. Columns T=1, T=2, T=4, T=8, and T=16 correspond to the number of threads used, while the last column shows the obtained speedup with 16 OpenMP threads (T=16) compared to one OpenMP thread (T=1). Note that the reduction in the execution time is not solely due to the introduction of multiple threads, as the improvements in the FFT routine used also have noticeable impact. This is most evident when comparing execution times of serial versions to OpenMP versions with one thread. Execution times and speedups of imag1dZ-th, real1dZ-th, imag2dXZ-th, and real2dXZ-th (not reported here) are similar to those of imag1dX-th, real1dX-th, imag2dXY-th, and real2dXY-th, respectively.}
\centering
\begin{tabular}{lcccccccc}
\hline
& Serial \cite{dbec2015} & T=1    & T=2    & T=4    & T=8    & T=16   & speedup \\
\hline
imag1dX-th & 9.1    & 7.1    & 4.7    & 3.4    & 2.9    & 2.8    & 2.5     \\
real1dX-th & 15.2   & 14.2   & 10.5   & 8.2    & 7.3    & 7.2    & 2.0     \\
\hline
imag2dXY-th & 13657  & 7314   & 4215   & 2159   & 1193   & 798    & 9.2     \\
real2dXY-th & 17281  & 11700  & 6417   & 3271   & 1730   & 1052   & 11.1    \\
\hline
imag3d-th & 16064  & 9353   & 5201   & 2734   & 1473   & 888    & 10.5    \\
real3d-th & 22611  & 17496  & 9434   & 4935   & 2602   & 1466   & 11.9    \\
\hline
\label{tab1}
\end{tabular}
\end{table}

The testing of OpenMP versions of programs DBEC-GP-OMP was performed with the number of threads varying from 1 to 16. Table~\ref{tab1} and Fig.~\ref{fig4} show the obtained absolute wall-clock times, speedups, and scaling efficiencies, as well as comparison with the previous serial version of programs \cite{dbec2015}. As we can see from the table, improvements in the FFT routine used already yield a speedup of 1.3 to 1.9 for single-threaded (T=1) 2d and 3d programs compared to the previous serial programs, and somewhat smaller speedup for 1d programs, 1.1 to 1.3. The use of additional threads brings about further speedup of 2 to 2.5 for 1d programs, and 9 to 12 for 2d and 3d programs. From Fig.~\ref{fig4} we see that for 1d programs, although speedup increases with the number of threads used, the efficiency decreases due to insufficient size of the problem, and one can achieve almost maximal value of speedup already with T=4 threads, while still keeping the efficiency around 50\%. We also see, as expected, that speedup and efficiency of 2d and 3d programs behave quite well as we increase the numbers of threads. In particular, we note that the efficiency is always above 60\%, making the use of all available CPU cores worthwhile.

\begin{figure}[!t]
\centering
\includegraphics[width=6.5cm]{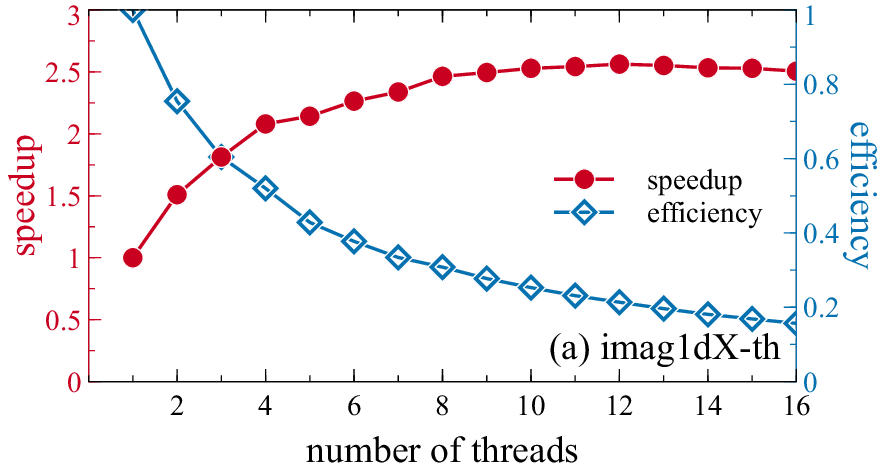}\hspace*{5mm}
\includegraphics[width=6.5cm]{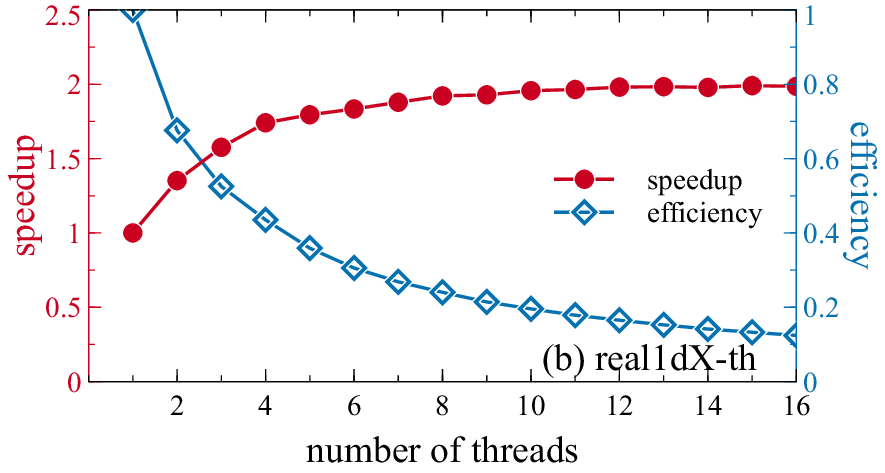}
\includegraphics[width=6.5cm]{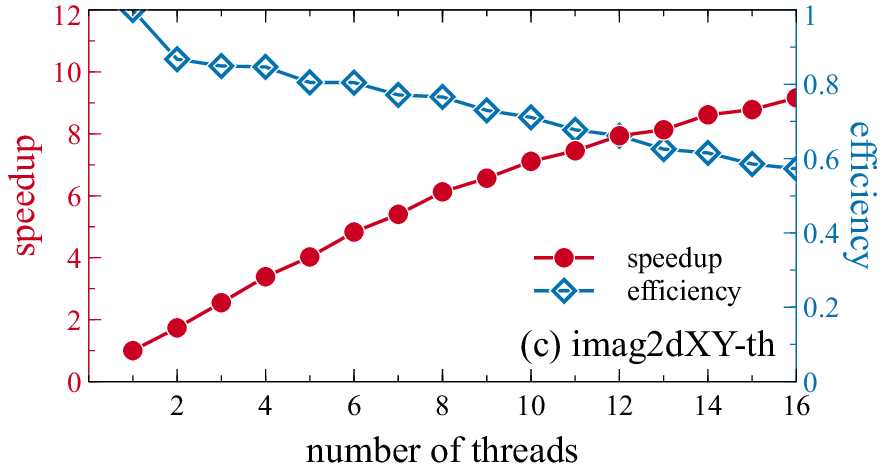}\hspace*{5mm}
\includegraphics[width=6.5cm]{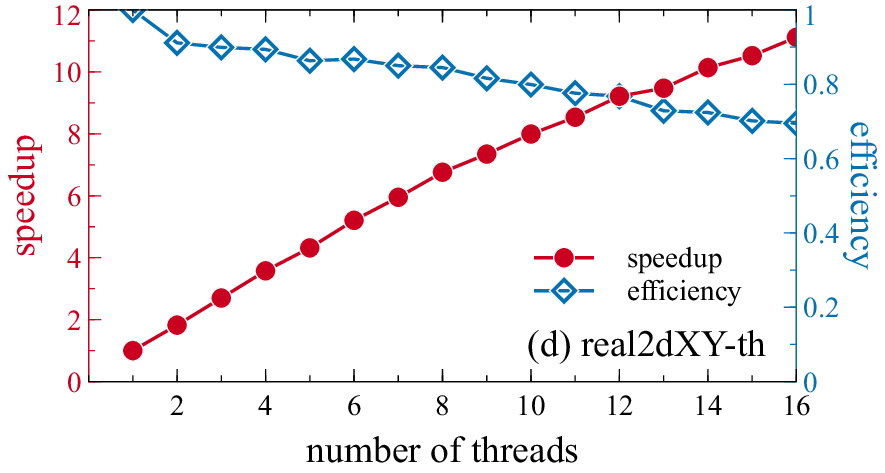}
\includegraphics[width=6.5cm]{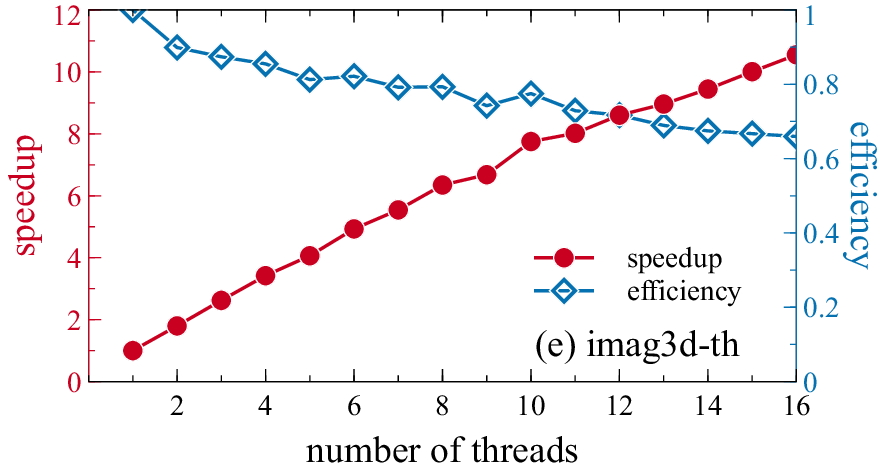}\hspace*{5mm}
\includegraphics[width=6.5cm]{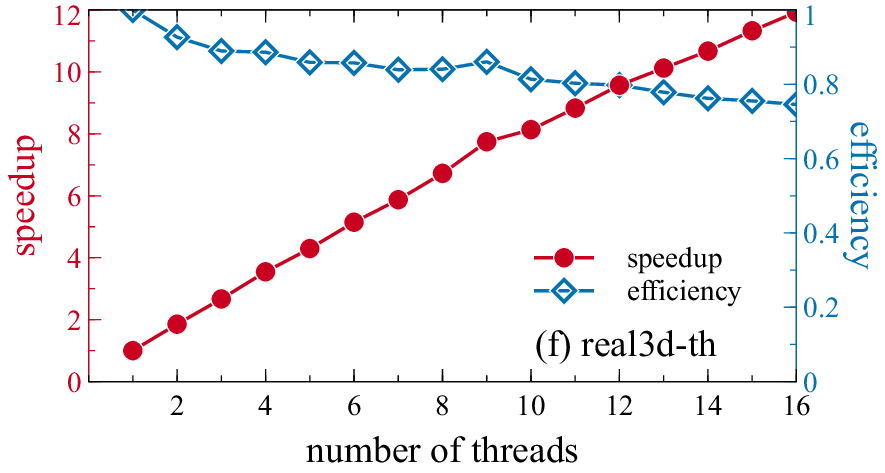}
\caption{Speedup in the execution time and scaling efficiency of DBEC-GP-OMP programs compared to single-threaded runs: (a) imag1dX-th, (b) real1dX-th, (c) imag2dXY-th, (d) real2dXY-th, (e) imag3d-th, (f) real3d-th. Scaling efficiency is calculated as a fraction of the obtained speedup compared to a theoretical maximum. Grid sizes used for testing are the same as in Table~\ref{tab1}. Speedups and efficiencies of imag1dZ-th, real1dZ-th, imag2dXZ-th, and real2dXZ-th (not reported here) are similar to those of imag1dX-th, real1dX-th, imag2dXY-th, and real2dXY-th, respectively.}
\label{fig4}
\end{figure}

\begin{table}[!b]
\caption{Wall-clock execution times of DBEC-GP-MPI programs compiled with mpicc compiler from OpenMPI implementation of MPI, backed by Intel's icc compiler, compared to the execution times of OpenMP (DBEC-GP-OMP) versions on a single-node (T=16, N=1). The execution times given here are for 1000 iterations (in seconds, excluding initialization and input/output operations, as reported by each program) with the grid size $480\times480\times500$. Columns N=4, N=8, N=16, N=24, and N=32 correspond to the number of cluster nodes used (each with T=16 threads), while the last column shows the obtained speedup with N=32 nodes compared to single-node runs.}
\centering
\begin{tabular}{lcccccccc}
\hline
           & OpenMP & N=4    & N=8    & N=16   & N=24   & N=32   & speedup \\
\hline
imag3d-mpi & 1124   & 653    & 352    & 167    & 128    & 96     & 11.5    \\
real3d-mpi & 2140   & 979    & 513    & 277    & 220    & 129    & 16.5    \\
\hline
\label{tab2}
\end{tabular}
\end{table}

For testing of MPI versions we have used a similar methodology to measure the strong scaling performance. For OpenMP/MPI programs DBEC-GP-MPI, the obtained wall-clock times are shown in Table~\ref{tab2}, together with the corresponding wall-clock times for the OpenMP programs DBEC-GP-OMP that served as a baseline to calculate speedups. The testing was done for varying number of cluster nodes, from 4 to 32, and the measured speedup ranged from 11 to 16.5. The corresponding graphs of speedups and efficiencies are shown in Fig.~\ref{fig5}, where we can see that the speedup grows linearly with the number of nodes used, while the efficiency remains mostly constant in the range between 40\% and 60\%, thus making the use of OpenMP/MPI programs highly advantageous for problems with large grid sizes.

\begin{figure}[!t]
	\centering
	\includegraphics[width=6.5cm]{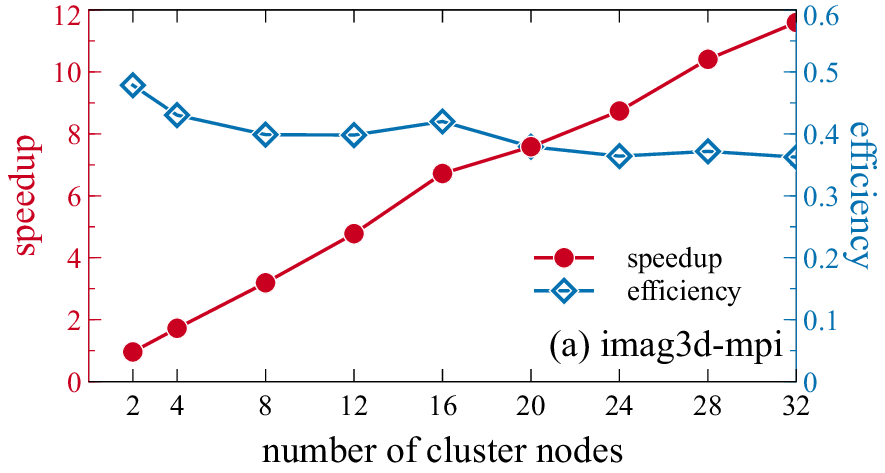}\hspace*{5mm}
	\includegraphics[width=6.5cm]{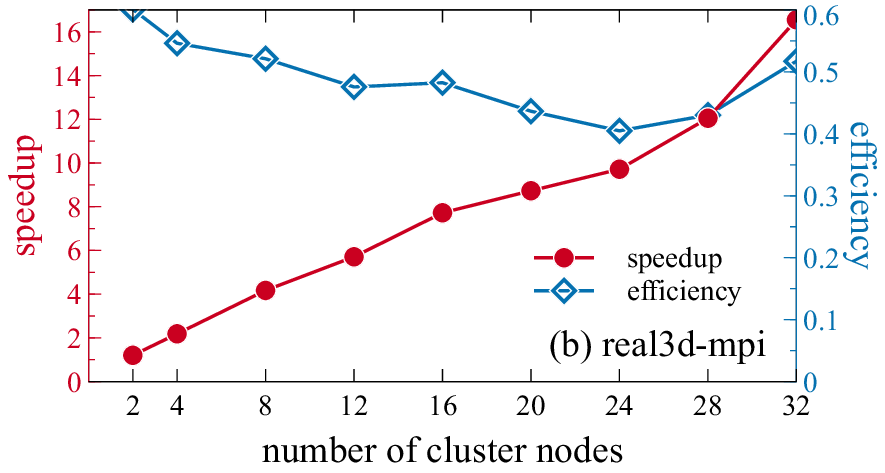}
	\caption{Speedup in the execution time and scaling efficiency of DBEC-GP-MPI programs compared to single-node OpenMP runs: (a) imag3d-mpi, (b) real3d-mpi. Scaling efficiency is calculated as a fraction of the obtained speedup compared to a theoretical maximum. Grid size used for testing is the same as in Table~\ref{tab2}.}
	\label{fig5}
\end{figure}

For CUDA/MPI programs DBEC-GP-MPI-CUDA we observe similar behavior in Table~\ref{tab3} and in Fig.~\ref{fig6}. The obtained speedup with N=32 nodes here ranges from 9 to 10, with the efficiency between 30\% and 40\%. While the efficiency is slightly lower than in the case of OpenMP/MPI programs, which could be expected due to a more complex memory hierarchy when dealing with the multi-GPU system distributed over many cluster nodes, the speedup still grows linearly and makes CUDA/MPI programs ideal choice for use on GPU-enabled computer clusters. Additional benefit of using these programs is their low CPU usage (up to one CPU core), allowing for the possibility that same cluster nodes are used for other CPU-intensive simulations.

\begin{table}[!ht]
\caption{Wall-clock execution times of DBEC-GP-MPI-CUDA programs compiled with Nvidia's nvcc compiler, with CUDA-aware OpenMPI implementation of MPI, backed by Intel's icc compiler, compared to the execution times of previous CUDA \cite{cuda2016} versions on a single-node with one GPU card (N=1). The execution times given here are for 1000 iterations (in seconds, excluding initialization and input/output operations, as reported by each program) with the grid size $480\times480\times250$. Columns N=4, N=8, N=16, N=24, and N=32 correspond to the number of cluster nodes used (each with one GPU card), while the last column shows the obtained speedup with N=32 nodes compared to single-node runs.}
\centering
\begin{tabular}{lcccccccc}
\hline
               & CUDA \cite{cuda2016}  & N=4    & N=8    & N=16   & N=24   & N=32   & speedup \\
\hline
imag3d-mpicuda & 579    & 447    & 212    & 103    & 71     & 61     & 9.5     \\
real3d-mpicuda & 800    & 619    & 295    & 142    & 96     & 80     & 9.9     \\
\hline
\label{tab3}
\end{tabular}
\end{table}

\begin{figure}[!ht]
	\centering
	\includegraphics[width=6.5cm]{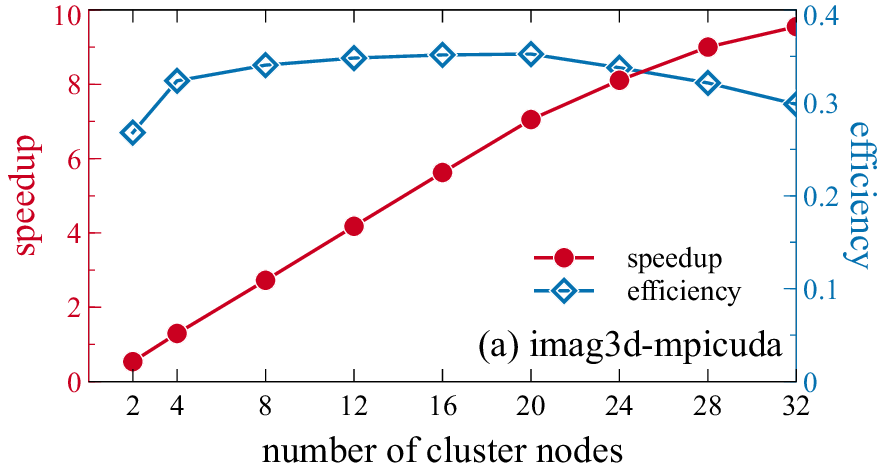}\hspace*{5mm}
	\includegraphics[width=6.5cm]{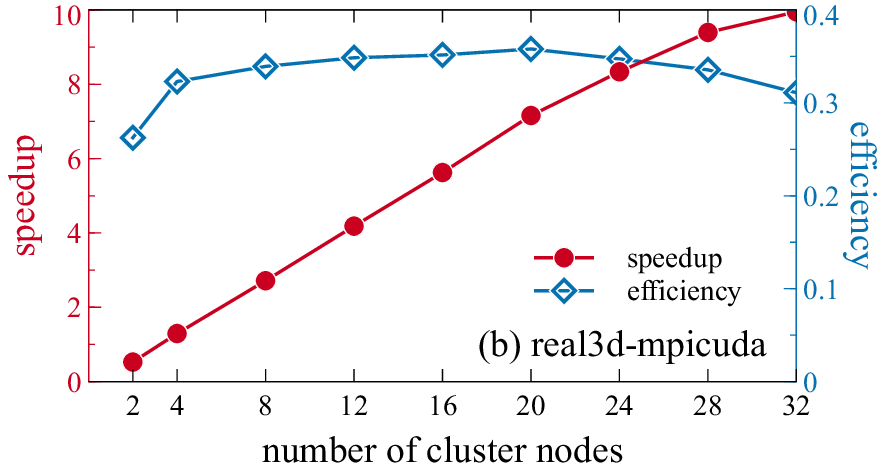}
	\caption{Speedup in the execution time and scaling efficiency of DBEC-GP-MPI-CUDA programs compared to single-node runs of previous CUDA programs \cite{cuda2016}: (a) imag3d-mpicuda, (b) real3d-mpicuda. Scaling efficiency is calculated as a fraction of the obtained speedup compared to a theoretical maximum. Grid size used for testing is the same as in Table~\ref{tab3}.}
	\label{fig6}
\end{figure}

The introduction of distributed transposes of data creates some overhead, which negatively impacts scaling efficiency. This is more evident in the CUDA/MPI version, as the transpose algorithm is inferior to the one provided by FFTW3. In our tests, both MPI versions of programs failed to achieve speedup on less than 4 nodes, due to the introduction of the transpose routines. We therefore recommend using MPI versions only on 4 or more cluster nodes.

The MPI versions are not only highly dependent on the configuration of the cluster, mainly on the speed of interconnect, but also on the distribution of processes and threads, NUMA configuration, etc. We recommend that users experiment with several different configurations to achieve the best performance. The results presented are obtained without extensive tuning, with the aim to show the base performance. 

Finally, we note that the best performance can be achieved by evenly distributing the workload among the MPI processes and OpenMP threads, and by using grid sizes which are optimal for FFT. In particular, the programs in DBEC-GP-OMP package have the best performance if NX, NY, and NZ are divisible by the number of OpenMP threads used. Similarly, for DBEC-GP-MPI programs the best performance is achieved if NX and NY are divisible by a product of the number of MPI processes and the number of OpenMP threads used. For DBEC-GP-MPI-CUDA programs, the best performance is achieved if NX and NY are divisible by a product of the number of MPI processes and the number of Streaming Multiprocessors (SM) in the GPU used. For all three packages, the best FFT performance is obtained if NX, NY and NZ can be expressed as $2^a 3^b 5^c 7^d 11^e 13^f$, where $e$ and $f$ are either 0 or 1, and the other exponents are non-negative integer numbers \cite{fftperf}.

\noindent\\
{\em Additional comments, restrictions, and unusual features:}
MPI programs require that grid size (controlled by input parameters NX, NY and NZ) can be evenly distributed between the processes, i.e., that NX and NY are divisible by the number of MPI processes. Since the data is never distributed along the NZ dimension, there is no such requirement on NZ. Programs will test if these conditions are met, and inform the user if not (by reporting an error). Additionally, MPI versions of CUDA programs require CUDA-aware MPI implementation. This allows the MPI runtime to directly access GPU memory pointers and avoid having to copy the data to main RAM. List of CUDA-aware MPI implementations can be found in Ref.~\cite{cudampi}.

\section*{Acknowledgements}
\noindent
V.~L., S.~\v S. and A.~B. acknowledge support by the Ministry of Education, Science, and Technological Development of the Republic of Serbia under projects ON171017, OI1611005, and III43007, as well as SCOPES project IZ74Z0-160453.
L.~E. Y.-S. acknowledges support by the FAPESP of Brazil under project 2012/21871-7 and 2014/16363-8.
P.~M. acknowledges support by the Science and Engineering Research Board, Department of Science and Technology, Government of India under project No.~EMR/2014/000644.
S.~K.~A. acknowledges support by the CNPq of Brazil under project 303280/2014-0, and by the FAPESP of Brazil under project 2012/00451-0.

\end{small}

\end{document}